\RequirePackage{fix-cm}

\documentclass[smallextended]{svjour3}       
\smartqed  
\usepackage{graphicx}
\usepackage[round]{natbib}
\usepackage{dcolumn}
\usepackage{bm}
\usepackage{physics}
\usepackage{bbm}
\usepackage{amsmath}
\usepackage{mhchem}
\usepackage{placeins}
 \usepackage{subcaption}
\usepackage{lipsum}
\usepackage{amssymb}
\usepackage{wrapfig}

\begin{document}

\title{A Reinforcement Learning approach for Quantum State Engineering
}


\author{Jelena Mackeprang   \and Durga  B Rao Dasari  \and  J\"org  Wrachtrup}

\authorrunning{Jelena Mackeprang   \and Durga  Dasari  \and  J\"org  Wrachtrup} 

\institute{J. Mackeprang \at
              3 Physikalisches Institut, Universit\"at Stuttgart, 70569 Stuttgart, Germany \\
              \email{jelena.mackeprang@pi3.uni-stuttgart.de}           
           \and
           D.  Dasari  \at
              3 Physikalisches Institut, Universit\"at Stuttgart, 70569 Stuttgart, Germany \\
               \email{d.dasari@pi3.uni-stuttgart.de}
               \and
              J.  Wrachtrup \at
              3 Physikalisches Institut, Universit\"at Stuttgart, 70569 Stuttgart, Germany \\
              Max Planck Institute for Solid State Research, 70569, Stuttgart, Germany
               }

\date{Received: date / Accepted: date}

\maketitle

\begin{abstract}
Machine learning (ML) has become an attractive tool in information processing, however few ML algorithms have been successfully applied in the quantum domain. We show here how classical reinforcement learning (RL) could be used as a tool for quantum state engineering (QSE). We employ a measurement based control for QSE where the action sequences are determined by the choice of the measurement basis and the reward through the fidelity of obtaining the target state. Our analysis clearly displays a learning feature in QSE, for example in preparing arbitrary two-qubit entangled states. It delivers successful action sequences, that generalise previously found human solutions from exact quantum dynamics. We provide a systematic algorithmic approach for using RL algorithms for quantum protocols that deal with non-trivial continuous state (parameter) space, and discuss on scaling of these approaches for preparation of arbitrarily large entangled (cluster) states.
 
\end{abstract}

\section*{Introduction}
\label{intro}
Current advancements in quantum technology have rendered the research field and its applications increasingly complex. As Machine Learning (ML 
) has been successfully applied to various classical problems as image recognition \citep{real_large-scale_2017} \citep{krizhevsky_imagenet_2012}, natural language processing \citep{lample_phrase-based_2018} and board games \citep{silver_general_2018}, its applicability to quantum problems has gained increasing interest. There are three main variants of ML: supervised, unsupervised and reinforcement learning (RL). The latter is based on an abstract agent interacting with its environment and receiving information in the form of a reward. It has already been used 
the optimisation of quantum control \citep{bukov_reinforcement_2018}. 
One key difference between RL and other types of machine learning is how information is gathered. For example in Supervised Learning, one starts with a large amount of labelled data, e.g. a set of input values $M$ of some unknown function $f: X \rightarrow Y$ and its respective set $N$ of output values or targets $y_{\mathrm{t}}$, how they are usually called, in order to predict the output of said function for the set $X \backslash M$ of input values not found in the original data set. 
In RL, however, the agent dynamically selects the data it wants to generate by selecting certain actions and observing the subsequent state transitions. This way it resembles the process of someone carefully designing an experiment to gain the desired information. 


Entangled quantum  states  are  crucial  resources  for  quantum optics  and  information  tasks such as precision sensing, quantum communication and computing. Construction and maintenance of  such  states  have  hence  been  a  long  term  challenge  involving  clever  use  of  interactions between physical systems  with ancillary degrees of freedom. One approach for preparing useful entangled states has been achieved through quantum measurements, where by manipulating the state of a sub-system via sequences of (projective) measurements or other operations on this sub-system, the rest of the system can be lead to the desired entangled state. Here, we dub this approach quantum state engineering (QSE).

In this work we use a RL approach to obtain optimal control sequences for a QSE problem, that generalise previous human solutions. \\

RL is divided into two abstract entities, the agent and the environment. The agent (the learner), is supposed to reach a certain goal by performing actions $a$ that change the state $s$ of the environment (which is defined by everything outside of the agent). Neither the agent nor the environment necessarily have to be physical, i.e. the agent does not have to be an actual physical manifestation of an intelligent entity, like a robot. \\ The entire learning process is usually divided into so called episodes, at the beginning of which the environment is reset to a starting state $s_0$. The episodes are in turn divided into discrete time steps $t$ (where the time does not necessarily have to correspond to physical reality), at which the agent selects an action to take. These actions end upon satisfying a certain criterion, e.g. when the maximum number of allowed time steps has passed or when the goal is achieved. 
At each time step the agent is given the state $s$ of the environment, chooses an action $a$ that changes the state and then receives feedback about the effectiveness of the performed action with regard to reaching its end goal in the form of a reward $r$. All rewards received after a time step $t$ accumulate to a return $G_{\mathrm{t}}$ defined by:

\begin{equation} \label{eq:return}
G_{\mathrm{t}} = \sum_{\mathrm{i}=0}^{\infty} \gamma^i \, r_{t+i+1}
\end{equation} 

The condition for most RL algorithms to work return is a sum over all future rewards, discounted by a factor $\gamma$ with $0 \leq \gamma \leq 1$ that controls the convergence.
After having repeated this procedure for a long enough time, the agent is supposed to have the ability to choose the appropriate action at each state it encounters such that the return at each time step is maximised. \citep{sutton_reinforcement_1998}\\

The necessary condition for most RL algorithms to work is that the problem can be formulated as a Markov Decision Process (MDP), i.e. the probability of the environment entering the state $\prime{s}$ and the agent receiving a reward $r$ must only depend on the current state $s$ and the current chosen action $a$. In other words: It should not matter how the environment transitioned to a state $\prime{s}$ at a time step $t$ and what the agent did beforehand. While this requirement is naturally met for classical problems, quantum dynamics is inherently non-Markovian, and the choice of the path traversed during a quantum evolution is a crucial information. Hence, in applying RL for quantum protocols one needs to carefully analyse the underlying problem to be studied.
\\

\section*{Modelling} 

Usually, the agent tries to predict so called optimal action-values $q(s,a)$ for the states $s$ and the actions $a$. The optimal action-value of a state $s$ and an action $a$ at the time step $t$ is defined as the maximum expected return after it for a state $s$ and for a selected action $a$. The equation governing all RL algorithms relates the optimal action-value $q(s,a)$ of a state $s$ and and action $a$ with the optimal action value of the successive state $s_{\mathrm{t+1}}$ over all possible actions and is called the Bellman optimality equation:  

\begin{equation}\label{eq:bellopt_q}
 q^{\text{*}}(s,a) = \mathbb{E}  \left[ r_{t+1}  + \gamma \, \underset{a'}{\mathrm{max}}\, q^{\text{*}}(s_{t+1},a') \,|\,  s_t = s,\,a_t = a\right]
 \end{equation}
  Here, $\mathbb{E}$ denotes an expectation value. \\
 
 We implement RL for a state space described by sets of continuous variables where the actions the agent can choose transfer the system from one state to the other. Our goal is for the state of the environment to reach a state that fulfils a predetermined criterion. The reward should therefore inform the agent whether or not the criterion has been satisfied.
 The fact that we are dealing with mixed quantum states constitutes a problem. Usually, when the goal is to lead the system described by a density matrix to a state that satisfies a certain condition, the RL state space $S$ has to consist of all independent entries of the entire density matrix in order for the Markov criterion to be fulfilled. This means that we deal with a exponentially growing, multi-dimensional continuous, unbounded, state space, whose optimal action-value function $q(s,a)$ has to be approximated. \\

One class of function approximators that have been shown to be able to master large, high dimensional, continuous input data are neural networks. \citep{mehta_high-bias_2018} One of their most prominent achievements was the classfication of images that were input in the form of raw pixel data into subcategories as cats and dogs.
 \citep{krizhevsky_imagenet_2012}.
Hence we use the Deep Q-Network \citep{mnih_human-level_2015} (DQN) and the Double Deep Q-Network \citep{hasselt_deep_2016} (DDQN) algorithm that combine classical RL with artifical neural networks (ANNs). In both algorithms, an ANN takes on the role of the agent's "brain" that approximates the optimal action-value-function $q(s,a)$ returning the action value $q(s,a)$ for any state $s$ represented by the set of continuous variables and any action $a$. 

At the beginning, the outputs $q(s,a)$ for the inputs $s$ and $a$ of the ANN will be random, but over time, as the agent gains experience by observing its received rewards, its internal parameters determining its output are adapted by taking advantage of equation \eqref{eq:bellopt_q}.
Here, the function $f: X \rightarrow Y$ to be approximated by the neural network is the action value function $q: S \cup A \rightarrow  Q $ where $Q$ is the space of all possible optimal action values. At each training step (of the entire learning algorithm), the agent "remembers" transition points $(s,a,s',r)$ after passing through a certain number of episodes, i.e. an observed state $s$, a selected action $a$, the observed subsequent state $s'$ and the received reward $r$, and stores them in a so called replay memory. After the agent has carried out a fixed number of episodes, the neural network (in the case of DQN) updates the previously described targets $y_{\mathrm{t}}(s,a)$ according to the following equation:

\begin{equation}\label{eq:DQNtargets} 
 y_{\mathrm{t}}(s,a) = \begin{cases} r &\mbox{if}\, s' \,\mbox{is terminal} \\ 
r+\gamma \,\underset{a'}{\mathrm{max}}\,Q(s',a') & \mbox{else}  \end{cases}   
\end{equation} 

This is repeated until the agent's performance is deemed satisfying. As one can see, the targets for the action value function as predicted by the neural network are updated using its current approximation also given by the network. The action $a'$ that maximises the action value for a given state $s$ is determined via the same function that also outputs the action value for said action $a'$ and for the state $s$ for the calculation of the new targets. The ever changing targets may lead to instability of the function approximation process. The DDQN algorithm tries to avoid this by introducing a second neural network, the target network, that is utilised to calculate the new targets. While the main network determines the maximising action $a'$, the target network gives the action value for said action and state. The network that the agent uses as a reference and whose weights are constantly updated is the main network. Meanwhile, the internal parameters of the target network are slowly adjusted to the ones of the main one over the course of the entire learning process.  \citep{hasselt_deep_2016} Then main network's targets are updated by:  

 \begin{equation}\label{eq:DDQNtargets} 
 y_{\mathrm{t}}(s,a) = \begin{cases} r, \\
 \qquad s' \,\mbox{terminal} \\ 
r+\gamma \,Q_{\mathrm{t}}(s',\underset{a'}{\mathrm{amax}}\,Q(s',a')), \\ \qquad \mbox{else}  \end{cases}  
\raisetag{\baselineskip}
\end{equation}

Due to few interactions between the agent and the environment at the beginning of the learning process, it is essential to make sure that it explores a large number of possibilities before determining an action-value-function to be optimal. Otherwise it could happen that the agent already performs well but has not found the optimal policy, i.e. generates a seemingly large return for each time step, but cannot find an even better strategy as it is stuck at a local maximum of the action value function with respect to the internal parameters of the neural networks. In order to avoid this problem a common used trick in the literature is to let the agent choose the action $a$ with the highest action value according to its own approximation for a given state $s$ at a time step $t$ with a probability $1-\epsilon$ and let it choose an action randomly with a probability $\epsilon$. This is called an epsilon-greedy policy. \citep{sutton_reinforcement_1998}
 We adapt this policy for our problem and linearly decrease $\epsilon$ at each step of the DQN/DDQN algorithm until it has reached a final predetermined minimum value $\epsilon_{\mathrm{min}}$.\\
 With this we show that the agent can find the same combinations as humans but can also come up with its own solutions. 

 \begin{figure}

        \includegraphics[width=0.75\textwidth]{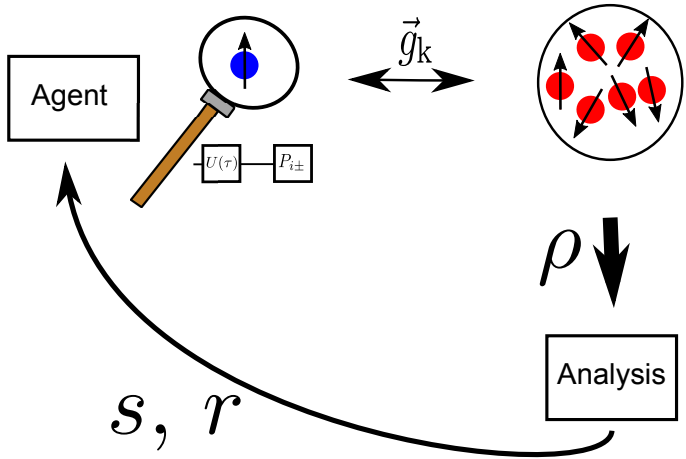}
       \caption{Depiction of our RL scheme. The agent influences the central spin's and therefore the nuclear spin's state by applying one of the projection operators (or none of them) after letting the entire system freely evolve for a time $\tau$. The spins' density matrix is then analysed and all necessary information is given to the agent who then decides over its next action, i.e. what projective measurement it performs, completing the circle.}
    \label{fig:Figure1}
    
\end{figure}

\section*{Quantum state engineering}
\label{sec:QSE}
Here we implement RL for a central spin model composed of a central spin-$1/2$ system coupled to a spin environment. Such a model can be physically realisable using solid state spins in diamond, where a single central electron spin, the Nitrogen Vacancy (NV) center, is coupled to an internal ${}^{13}C$ nuclear spin bath \citep{doherty_nitrogen-vacancy_2013}  \citep{wrachtrup}  . The central spin $S$ is a spin one ($S = 1$) system of which we choose a two-level subspace, and the nuclear spins ($I$) are spin-half $I=1/2$ particles, each with a two-level Hilbert space. The nuclear spins that constitute the spin-bath, are assumed to be weakly coupled and hence non-interacting for the present study. \\

With the magnetic field aligned along the central spin axis (say NV $z$-axis) the Hamiltonian that determines the dynamics is given by  \citep{greiner_purification_2017}:

\begin{equation}
H = S^{(z)} \otimes \sum_{\mathrm{k}}  \vec{g}_{\mathrm{k}} (r) \cdot \vec{I}_{\mathrm{k}} + \omega \sum_{\mathrm{k}} \mathbbm{1} \otimes I_{\mathrm{k}}^{(z)}
\label{eq:Ham}
\end{equation}

Here, $g_{\mathrm{k}}$ stands for the strength of the dipolar coupling between the NV (central) spin and the $k$-th nuclear spin, dependent on the spatial separation $r$ between these spins, $S^{(z)}$ and $I_{\mathrm{k}}^{(z)}$ for the spin components along the $z$-axis of the system and $\vec{I}_{\mathrm{k}}$  for the spin components along the three axes $x$, $y$ and $z$ of the k-th nuclear spin. The identity operator is denoted by $\mathbbm{1}$

Because of the random spatial location of the spins with respect to the NVC, the couplings become inhomogeneous such that the bath has neither conserved quantities nor preferred symmetries. The nuclear spins precess under the external field $\omega$ (the nuclear Zeeman term) that is acting along the $z$-direction and is assumed to be uniform. The dynamics generated by the Hamiltonian \eqref{eq:Ham} can be exactly solved. \citep{greiner_purification_2017}. 

The time evolution governed by it can be determined from the Schr\"odinger equation and is given by:

\begin{equation} \label{eq:timeev}
\rho'(t+\tau) = U(\tau) \, \rho(t)\, U^\dagger(\tau)
\end{equation}

Where $U(\tau) = \exp(-i\,(t-\tau) H)$ and $\rho'(t+\tau)$ is the time-evolved density matrix of the total system after a time interval $\tau$ with:

\begin{equation}\label{eq:U}
U(\tau) = \exp(-iH\tau)
\end{equation}

%
%
%

 To study the central spin model within the paradigm of reinforcement learning, we determine the agent to be a virtual experimentalist combining projective measurements via optical read out and time evolution determined by equation \eqref{eq:U} over fixed time intervals $\tau$. Therefore, the environment is everything outside of the agent, i.e. the entire physical system including the central spin and the nuclear spins. As the problem has to be formulated as a Markov Decision Process, the states are then best described by the independent entries of the density matrix $\rho$ describing all spins.
 At each step of an episode the agent lets the system freely evolve for a predetermined time $\tau$ and then performs either of the following actions:

\begin{enumerate}
\item{project the central spin onto the state $\ket{z+}$ by \newline applying the projection operator \newline $P_{z+} = \ket{z+}\bra{z+} \otimes \prod_{\mathrm{k}=1} \mathbbm{1} $}

\item{project the central spin onto the state $\ket{z-}$ by applying the Projection Operator $P_{z-} = \ket{z-}\bra{z-} \otimes  \prod_{\mathrm{k}=1} \mathbbm{1}$}

\item{project the central spin onto the state $\ket{x+} =1/\sqrt2\left( \ket{z+} + \ket{z-} \right)$ by applying the projection operator $P_{x+} = \ket{x+}\bra{x+} \otimes \prod_{\mathrm{k}=1} \mathbbm{1} $}

\item{project the central spin onto the state $\ket{x-} = 1/\sqrt2\left( \ket{z+} - \ket{z-} \right)$ by applying the Projection Operator $P_{x-} = \ket{x-}\bra{x-} \otimes \prod_{\mathrm{k}=1} \mathbbm{1} $}

\item{project the central spin onto the state $\ket{y+} =1/\sqrt2\left( \ket{z+} + i \ket{z-} \right)$ by applying the Projection Operator $P_{y+} = \ket{y+}\bra{y+} \otimes \prod_{\mathrm{k}=1} \mathbbm{1} $}

\item{project the central spin onto the state $\ket{y-} =1/\sqrt2\left( \ket{z+} - i \ket{z-} \right)$ by applying the Projection Operator $P_{y-} = \ket{y-}\bra{y-} \otimes \prod_{\mathrm{k}=1} \mathbbm{1}$}

\item do nothing,i.e. apply the identity operator $\mathbbm{1}$ to $\rho$.

\end{enumerate}

As the projective measurements are non unitary transformations, the time-evolved state  $\rho(t+\tau)$ must be appropriately normalised after every operation, for it to represent a valid density matrix. The density matrix $\rho_{t+\tau}$ at the time $t+\tau$ is derived from $\rho'_{t+\tau}$ following \eqref{eq:timeev} and by applying one of the projection operators $P_{i\pm}$  listed above (or none of them) and then normalised:  
\begin{equation}\label{eq:norm}
\rho_{t+\tau} = \frac{P_{i\pm} \, \rho'_{t+\tau} \, P_{i\pm}^{\dag}}{\mathrm{tr}(P_{i\pm} \, \rho'_{t+\tau} \, P_{i\pm}^{\dag})}
\end{equation}

During each episode the agent takes a maximum number $n_{\mathrm{e}}$ of steps. 
It terminates and a positive reward $r_+$ is given if the partial trace of density matrix $\rho(t+\tau)$ over the Hilbert space of the central spin, denoted by $\rho_I$, following the time evolution and projection, has an overlap with the target state $\ket{\Psi}$, i.e. the fidelity $F$, that exceeds a certain threshold $\theta$, for example $\theta = 0.99$. We define the fidelity $F(\sigma,\rho)$ of two density matrices $\sigma$ and $\rho$ as followed:

\begin{equation}\label{eq:fidelity}
F(\sigma,\rho) = \mathrm{tr}\sqrt{\sqrt{\rho}\sigma\sqrt{\rho}}.
\end{equation} 

As long as the fidelity is not greater than $\theta$ and the agent hasn't performed the maximum number of actions, a negative reward $r_-$ is given.
This way and by using all independent entries of the density matrix as the state representation we make sure that the reward received at some time $t$ and the subsequent state $s'$ only depends on the current chosen action $a$ (one of the projection operators or just the identity matrix) and the current state $s$, thereby fulfilling the Markov condition. The entire protocol is summarised by figure \ref{fig:Figure1}.\\

 
 \section{Construction of the Bell states} \label{sec:BSE}

\subsection*{fixed central spin starting state}
We consider the simplest case, i.e. a spin-bath composed of two nuclear spins. 
Therefore, the state of the environment is best described by a three-spin density matrix with $64$ complex entries of which $35$ are independent. Those 35 independent entries are then divided into their real and imaginary parts, resulting in an input of 70 continuous variables. 

At first, the environment is initialised as the starting state $s_0$ belonging to $\rho_0$ defined in equation \eqref{eq:StartState} at the beginning of each episode , i.e. the central spin is initialised as the $\ket{x+}$ - state following the promising results of the experiments reported by Johannes Greiner et. al \citep{greiner_purification_2017} where the central spin of the previously described system is always brought into a superposition state before any projective measurements are applied and they manage to purify the spin-bath with the help of a sequence of projective measurements and time evolutions, just as the agent is supposed to find. The nuclear spins are initialised in a completely mixed state:

 \begin{align} \label{eq:StartState}
    \rho_0 = \ket{x+}\bra{x+} \otimes \left(\begin{matrix} 
\frac{1}{2}& 0\\
0& \frac{1}{2} 
\end{matrix}\right)\otimes \left(\begin{matrix} 
\frac{1}{2}& 0\\
0& \frac{1}{2} 
\end{matrix}\right) 
 \end{align}

The density matrix of the nuclear spins $I_1$ and $I_2$ is written in the $\ket{m_\mathrm{s}= \pm 1} = \ket{z\pm} $ basis.

As mentioned above, each episode consists of a fixed number of repetitions and terminates when a maximum overlap of the nuclear spins' density matrix $\rho_I$ with the target state is achieved, or when all the $n_{\mathrm{e}}$ actions have been performed. We choose the four maximally entangled Bell states $\ket{\Phi^\pm}$ and $\ket{\Psi^ \pm}$  as our target states:

\begin{equation}
\ket{\Phi^\pm} = 1/\sqrt2 \, \left(\ket{z+}\ket{z+}\pm\ket{z-}\ket{z-}\right)
\end{equation}

\begin{equation}
\ket{\Psi^ \pm} = 1/\sqrt2 \, \left(\ket{z+}\ket{z-}\pm\ket{z-}\ket{z+}\right)
\end{equation}

The system is simulated with the parameters from equation \eqref{eq:Ham}
set to $\omega = 1/2$ and $\vec{g}_{\mathrm{k}} = (1,0,0)$ $\forall \, \mathrm{k}$ in relative units, i.e. the nuclear spins form a linear chain of spins arranged along the $x-$direction of the NVC. The time $\tau$ for which the agent lets the system freely evolve is set to $\tau =1$ relative to $\omega$. The reward and the indicator "$\mathrm{done}$" about whether or not an episode has terminated are defined by equation \eqref{eq:reward}, in which $m$ stands for the number of steps taken and $P_{i\pm}$ and $\rho'_{t+1}$ are defined in equation \eqref{eq:norm}. As the denominator in equation \eqref{eq:norm} can become too small for a computer to handle, some sort of punishment has to be introduced if the agent chooses a projective measurement that results in a computing error due to a denominator that is too close to zero. Receiving this reward $r_{\mathrm{fatal}}$ also results in the end of the episode. Summing all up, the reward $r$ is defined as:

\begin{equation} \label{eq:reward}
(r,\mathrm{done}) =  \begin{cases}
(r_+,\mathrm{1})&  \,F\left(\rho_I,\ket{\Psi}\bra{\Psi}\right)>\theta\\
(r_-,\text{0})& \,F\left(\rho_I,\ket{\Psi}\bra{\Psi}\right)<\theta \,\land\, m_{\mathrm{e}}<n_{\mathrm{e}}\\
(r_-,\text{1})& \,F\left(\rho_I,\ket{\Psi}\bra{\Psi}\right)<\theta \,\land\, m_{\mathrm{e}}=n_{\mathrm{e}}  

\\
(r_{\mathrm{fatal}},\mathrm{1}) & \mathrm{tr}\left(P_{i\pm} \, \rho'_{t+1} \, P_{i\pm}^{\dag}\right) \approx 0 

\end{cases}
\end{equation}

Here $m_\mathrm{e}$ stands for the number of steps taken and $n_\mathrm{e}$ for the maximum number of steps per episode.
For all cases mentioned in the following, $\theta$ is set to 0.99, $r$ to 10, and $r_-$ to -1.

The agent carries out a constant, sufficiently large number of episodes at each training step. As described above, it remembers the transitions and stores them in a replay memory. We have implemented an $\epsilon$- greedy policy with a linearly decaying $\epsilon$. To illustrate the agent's improving performance over time, figure \ref{fig:PsiMinusprogress} shows the average undiscounted sum of all rewards received during the episodes at each training step in the case of the Bell state $\ket{\Psi^-}$ where the system starts out with a density matrix given by equation \eqref{eq:StartState}:

\begin{figure}
\includegraphics[width=0.75\textwidth]{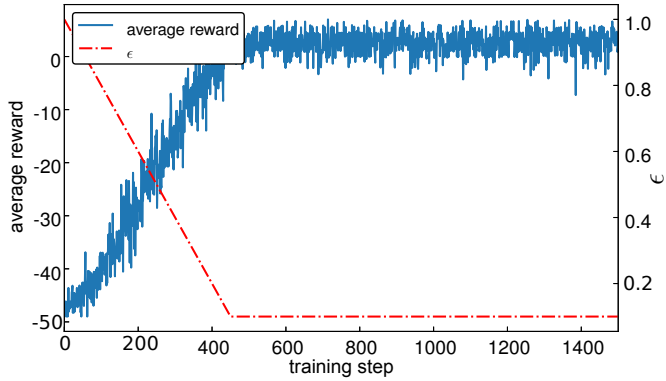}
\caption{Average over the total sum of all received rewards during the fixed number of episodes completed at each training step of the DQN algorithm for the target state $\ket{\Psi^{-}}$. The dashed line indicates the linearly decaying $\epsilon$. ($\epsilon$-greedy policy, see section Modelling)When it is small enough, it stops decaying and stays constant at $\epsilon = 0.1$. The maximum number of steps per episode $n_\mathrm{e}$ is set to 50 and $r_{\mathrm{fatal} } = -51$. At the beginning of each episode the central spin is initialised in the state $\ket{x+}$. Initially, the agent is rarely successful in achieving its goal but over time, its progress improves until it stays stagnant.}
\label{fig:PsiMinusprogress}
\end{figure}
 
We show the same for the target states $\ket{\Phi^-}$, $\ket{\Phi^+}$ and $\ket{\Psi^+}$ :

\begin{figure}
\includegraphics[width=0.75\textwidth]{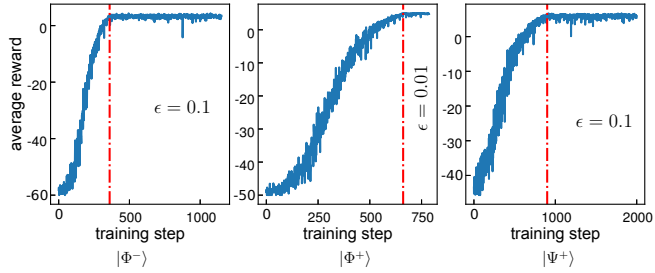}
\caption{Average over the total sum of all received rewards during the fixed number of episodes completed at each training step of the DQN algorithm for different target states, starting state: $\ket{x+}$. On the dashed red line's right side the linearly decaying $\epsilon$ has reached its minimum value and stays constant. For $\ket{\Phi^+}$ and $\ket{\Psi^+}$ $n_\mathrm{e}$ is equal to 50 and $r_{\mathrm{fatal}}$ to -51, for $\ket{\Phi^-}$ $n_\mathrm{e}$ is equal to 60 and $r_{\mathrm{fatal}}$ to -61. As for $\ket{\Psi^-}$ the agent on average receives a growing number of rewards per episode over time.}
\label{fig:PsiMinusprogressALL}

\end{figure}
\FloatBarrier

To make sure that it has found a policy superior to one where it randomly chooses an action at any time step, we compare the agent's performance at a small $\epsilon$ value to its behaviour at $\epsilon = 1$, i.e. when acting according to a completely random policy. 

\FloatBarrier

\begin{figure}
 
\includegraphics[width=0.75\textwidth]{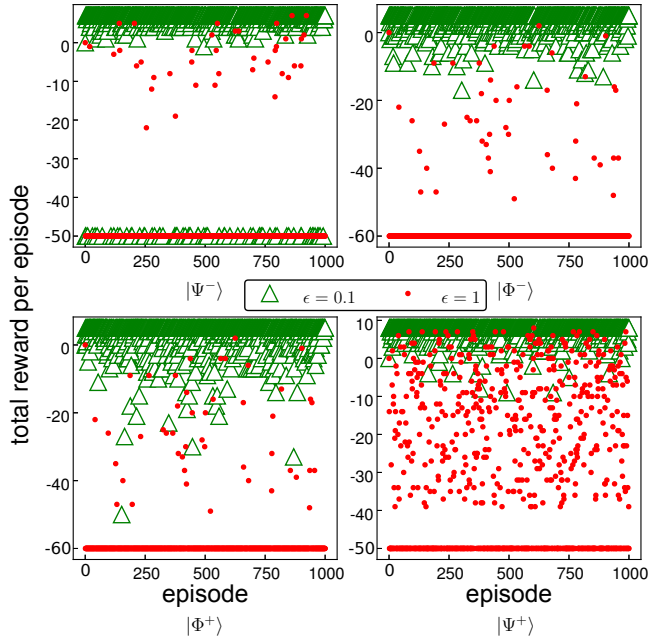}

\caption{Comparison of a randomly acting agent ($\epsilon = 1$, red dots) to a trained one ($\epsilon = 0.1$, green triangles) for different target states. Each point or triangle represents the total sum of received rewards during an episode. The agent receives a reward of 10 and terminates the episode if a fidelity of 0.99 with the target state is reached and receives one of -1 if the fidelity is below this threshold, i.e. the higher the total reward per episode, the fewer steps the agent needed to construct the desired target state.  The trained agents perform better than the ones acting according to a random policy.}

\label{fig:TFcomp}
\end{figure}

\begin{table*}
\caption[Sequences of time evolutions and subsequent projective measurements that lead to the different target states]{Sequences of time evolutions and subsequent projective measurements that lead to the four Bell states. Here, $\tau = 1$ (relative units) and $P_{i\pm}$ are the projection operators defined above. The central spin is initialised in the $\ket{x+}$ state and the fidelity is taken with respect to the target state. The last column lists the success rates of the respective sequences.}
 
\begin{tabular}{c|c|c|c}
 
   & Sequence & fidelity $\cdot \, 10^2$ & success rate $[\%]$  \\ \hline
$\ket{\Phi^+}$ & $U(2\,\tau)$-$P_{x+}$-$U(\tau)$-$P_{x+}$-$U(\tau)$-$P_{x-}$-$U(2\,\tau)$-$P_{x+}$-$U(\tau)$-$P_{x+}$&99.673  & 2.811 \\\hline 
 $\ket{\Phi^-}$ & $U(2\, \tau)$-$P_{x+}$-$U(\tau)$-$P_{x+}$-$U(\tau)$-$P_{x-}$-$U(\tau)$-$P_{y-}$-$U(\tau)$-$P_{x-}$-$U(\tau)$-$P_{x-}$&99.803 &0.998\\\hline 
 $\ket{\Psi^+}$&$U(\tau)$-$P_{x+}$-$U(2\,\tau)$-$P_{x+}$-$U(\tau)$-$P_{x+}$-$U(\tau)$-$P_{x-}$& 100.000 & 20.313 \\\hline  
 $\ket{\Psi^-}$&$U(2\,\tau)$-$P_{x+}$-$U(\tau)$-$P_{x+}$-$U(\tau)$-$P_{x+}$-$U(\tau)$-$P_{x+}$-$U(\tau)$-$P_{x+}$&99.454&25.275 \\\hline 
  $\ket{\Psi^-}$&$U(2\,\tau)$-$P_{y-}$-$U(\tau)$-$P_{y-}$-$U(\tau)$-$P_{y-}$-$U(\tau)$-$P_{y-}$-$U(\tau)$-$P_{y-}$ & 99.160 & 12.713   \\\hline 
  $\ket{\Psi^-}$& $U(\tau)$-$P_{x+}$-$U(\tau)$-$P_{y-}$-$U(\tau)$-$P_{y-}$-$U(\tau)$-$P_{y-}$-$U(\tau)$-$P_{y-}$ & 99.184 & 12.707  \\ 
 
\end{tabular}

\label{tab:successfulsequ}
\end{table*}

\begin{figure}[t!]
 
\includegraphics[width=0.75\textwidth]{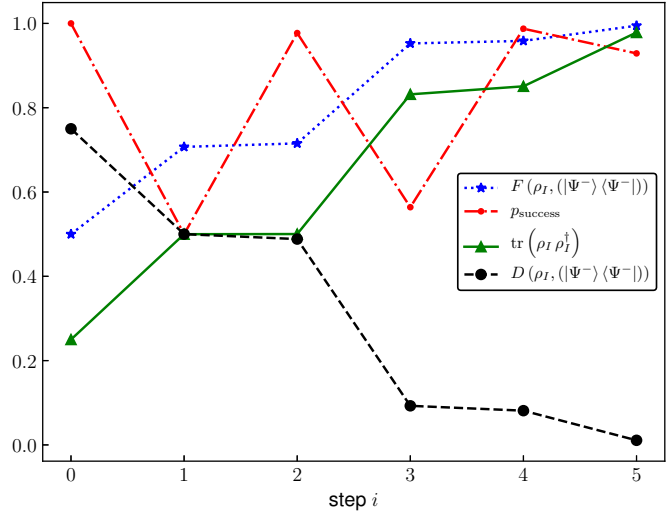}
\caption{Purity of the nuclear spin state, the fidelity and trace distance of it and the target state $\ket{\Psi^-}$ and the success probability of the projective measurement at each step of the sequence $U(2\,\tau)$-$P_{x+}$-$U(\tau)$-$P_{x+}$-$U(\tau)$-$P_{x+}$-$U(\tau)$-$P_{x+}$-$U(\tau)$-$P_{x+}$, that the agent has found. The central spin is initialised in the state $\ket{x+}$ and the nuclear spins in a completely mixed state.}
\label{fig:TFsequencePsiMinus}

\end{figure}

Table \ref{tab:successfulsequ} lists the sequences with the highest success rate the agent has found for the respective target states, where the success rates are the products of the success probabilities for each central spin read out, i.e. the denominator in equation \eqref{eq:norm}. Their most important aspects, i.e. the purity $\mathrm{tr}\left( \rho_I\rho_I^\dag\right)$ of the nuclear spin state, the fidelity $F\left(\rho_I,\ket{\Phi^+}\bra{\Phi^+}\right)$ of it and the desired Bell state, their trace distance and the success probability $\mathrm{tr}\left(P_{i\pm}\rho P_{i\pm}^\dag\right)$ of the projective measurement (or lack of it) at each step $\mathrm{i}$ are also visualised in figure \ref{fig:TFsequencePsiMinus} at the example of the first sequence in table \ref{tab:successfulsequ} for the target state $\ket{\Psi^-}$. Here, the trace distance of two density matrices $\rho$ and $\sigma$ is defined as:

\begin{equation}
D(\rho,\sigma) = \left(\sqrt{\left(\rho -\sigma\right) \, \left(\rho -\sigma\right)^{\dagger}}\right).
    \label{eq:trdst} 
\end{equation}

\begin{figure}[t!]
 
\includegraphics[width=0.75\textwidth]{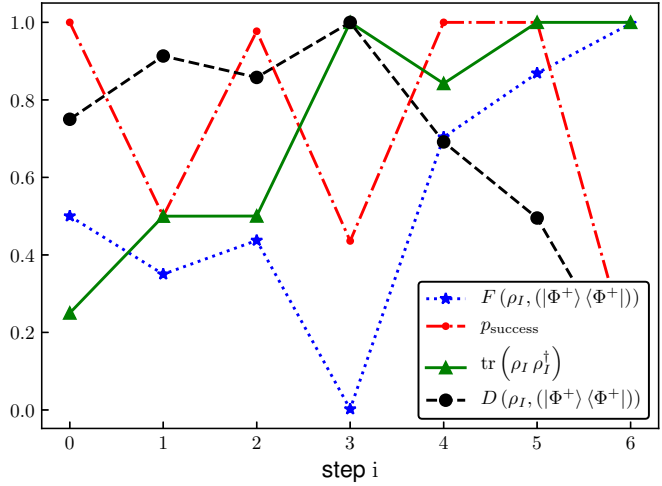}
\caption{Purity of the nuclear spin state, the fidelity and trace distance of it and the target state $\ket{\Phi^+}$ and the success probability of the projective measurement at each step of the sequence $U(2 \tau)$-$P_{x+}$-$U(\tau)$-$P_{x+}$-$U(\tau)$-$P_{x-}$-$U(2\tau)$-$P_{x+}$-$U(\tau)$-$P_{x+}$, that the agent has found. The central spin is initialised in the state $\ket{x+}$ and the nuclear spins in a completely mixed state. At each step the agent lets the system freely evolve for a time $\tau$ = 1 and then performs one of the projective measurements or none. The agent chose the latter option at the beginning of this sequence so the first two steps are summarised by $U(2 \tau)$-$P_{x+}$ and the success probability of the first is 1.}
\label{fig:TFsequencePhiPlus}

\end{figure}

Figure \ref{fig:TFsequencePsiMinus} shows a monotonically decreasing trace distance and a monotonically increasing fidelity which begs the question why the reward could not have been simply defined as the fidelity of the nuclear spin state with the target state at each step of the episode. The reason why this does not work can be seen at the example of one sequence the agent has found for the target state $\ket{\Phi^+}$ in figure \ref{fig:TFsequencePhiPlus}.

The data plotted in figure \ref{fig:TFsequencePhiPlus} clearly demonstrates that the reward must be defined as it is in equation \eqref{eq:reward}, i.e. only returning a positive value if the fidelity has exceeded the threshold $\theta$, since the fidelity does not increase continuously with the episode step $\mathrm{i}$. On the contrary,it plummets to nearly zero at the fourth step, i.e. an increasing or decreasing fidelity does not give the agent any information how to reach its end goal. \\

\begin{figure*}
\centering
 \includegraphics[width=0.75\textwidth]{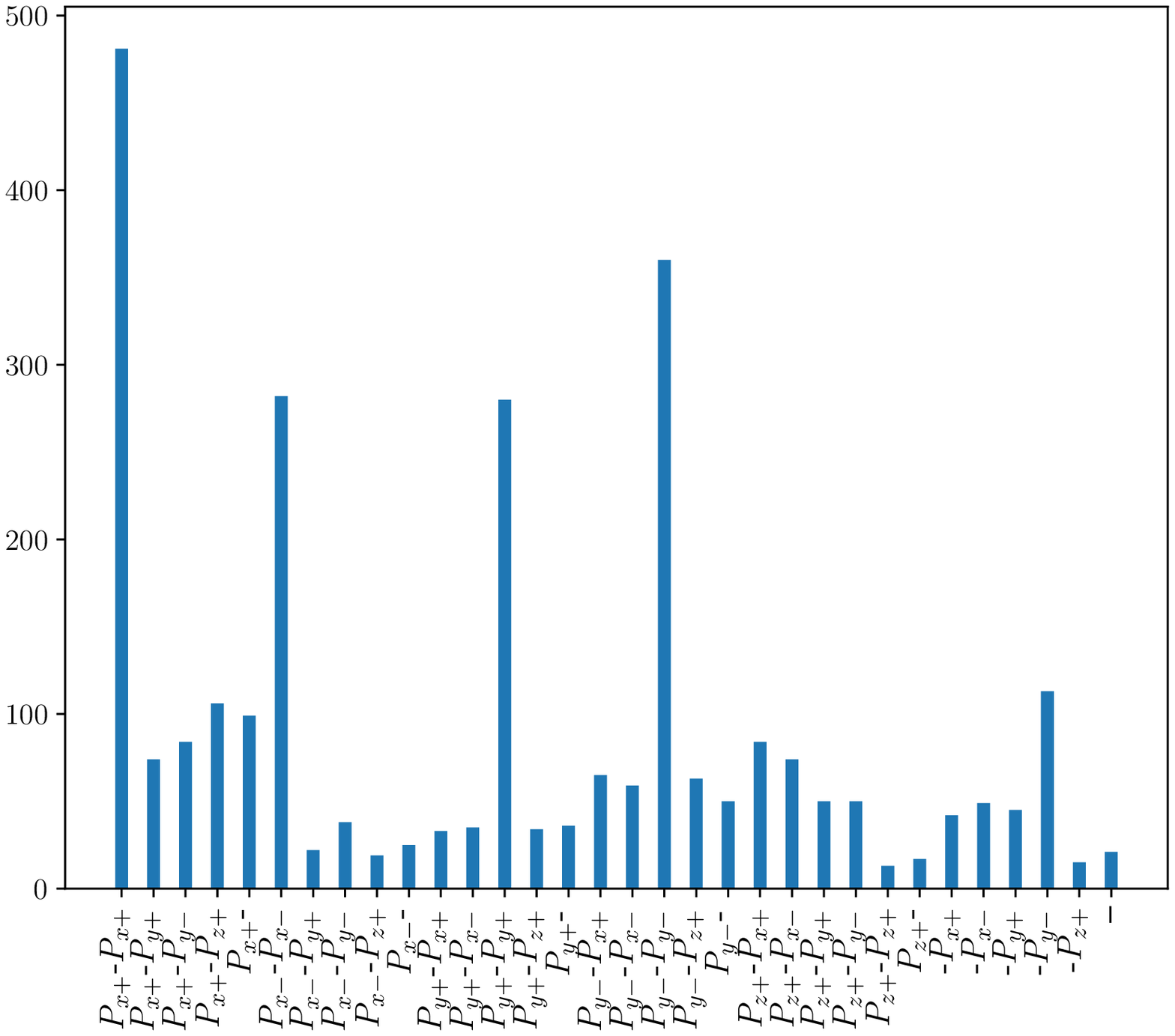}
    \caption{Counts of appearances of action combinations in the unique, successful sequences leading to $\ket{\Psi^-}$ taken from a total set of 3000 episodes. The dash stands for "do nothing" and the last bar corresponds to the agent choosing this option twice.}
    \label{fig:HistPsiMinus}
\end{figure*}

For the case of $\ket{\Psi^-}$ the agent has found the same sequence as Greiner et al. \citep{greiner_purification_2017}  where they showed that for a linear chain containing an even number of spins arranged along the $x-$axis of the system, the steady state of the nuclear spin ensemble evolves to

\begin{equation}\label{eq:evolved}
\ket{\Psi} = \otimes_{\mathrm{i=2}}^{\mathrm{N}} \ket{\Psi^-}
\end{equation}

when repeatedly applying $P_{x+}$ after the ensemble has freely evolved for a fixed time interval $\tau$. 
(N is the total number of spins divided by two.) However, the agent has also found many other sequences consisting of projections onto superposition subspaces, i.e. the $\ket{x\pm}$ and $\ket{y\pm}$ subspaces, as one can see in table \ref{tab:successfulsequ}. To further discuss this, the number of appearances of action combinations over the course of successful and unique episodes taken from a total set of 3000 episodes are illustrated in the histogram \ref{fig:HistPsiMinus}. 

Interestingly, the agent very much prefers repeating the same action over choosing another one except for the $P_{z+}$ projection and the option "do nothing". Apparently, repeatedly selecting either $P_{x+}$, $P_{x-}$, $P_{y+}$ or $P_{y-}$ during a sequence leads to $\ket{\Psi^-}$. This means that the agent has found other sequences besides the one described by Greiner et al., i.e. it generalised the solutions found by them.

\subsection*{random (pure) central spin starting state} 

Apart from initialising the environment in a definite state $s_0$ whose corresponding density matrix is  defined by equation \eqref{eq:StartState}, we also let the agent search for combinations of projective measurements when the state of the central spin is initialised in a random pure state $\ket{\Psi}$ at the beginning of each episode:

 \begin{align} \label{eq:RandomStartState}
    \rho_0 = \ket{\Psi}\bra{\Psi} \otimes \left(\begin{matrix} 
\frac{1}{2}& 0\\
0& \frac{1}{2} 
\end{matrix}\right)\otimes \left(\begin{matrix} 
\frac{1}{2}& 0\\
0& \frac{1}{2} 
\end{matrix}\right) 
 \end{align}

In the following we will only discuss the result for the target state $\ket{\Psi^-}$.
As the randomness leads to a larger variety of states the agent can encounter and therefore to increased noise, we utilise the Double DQN algorithm for more stable learning. We simulate the system with the same $\vec{g}_k$ and $\omega$ as before, but $\tau$ is now set to $2$ since the agent has frequently chosen to let the system evolve freely for the time $\tau = 2$ during the last step. (This is possible by selecting the action "do nothing".) The maximum number of steps per episode $n_{\mathrm{e}}$ is chosen to be 50 and $r_{\mathrm{fatal}}$ is set to -51. \\

\begin{figure}[t!]
    
\includegraphics[width=0.75\textwidth]{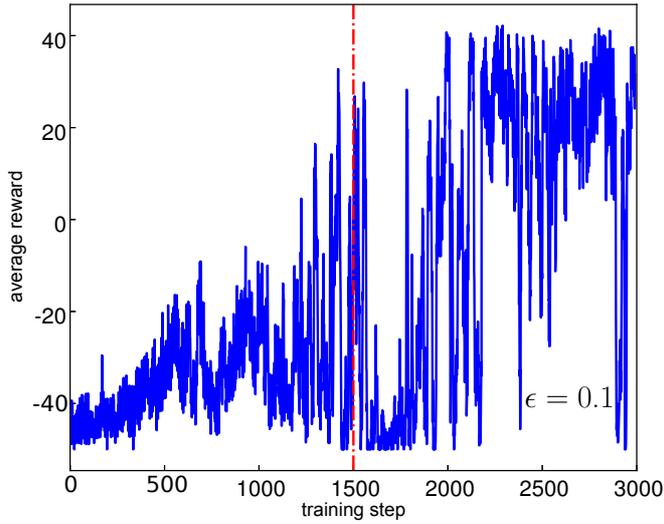} 
        \caption{Average over the total sum of all received rewards during the completed episodes at each training step of the Double DQN algorithm for the target state $\ket{\Psi^-}$. Instead of initialising the central spin in a fixed state, it is reset to a random pure state at the beginning of each episode.The agent's performance seems to increase, but its progress is highly unstable as indicated by the sudden dips and spikes of the learning curve.}
        \label{fig:PTprogress}
\end{figure}

The agent's progress is shown in figure \ref{fig:PTprogress}. It seems to be unstable which may be due to over-fitting of the under-laying neural networks. \citep{mehta_high-bias_2018}

\begin{figure}[t!]
    \centering

        \includegraphics[width=0.75\textwidth]{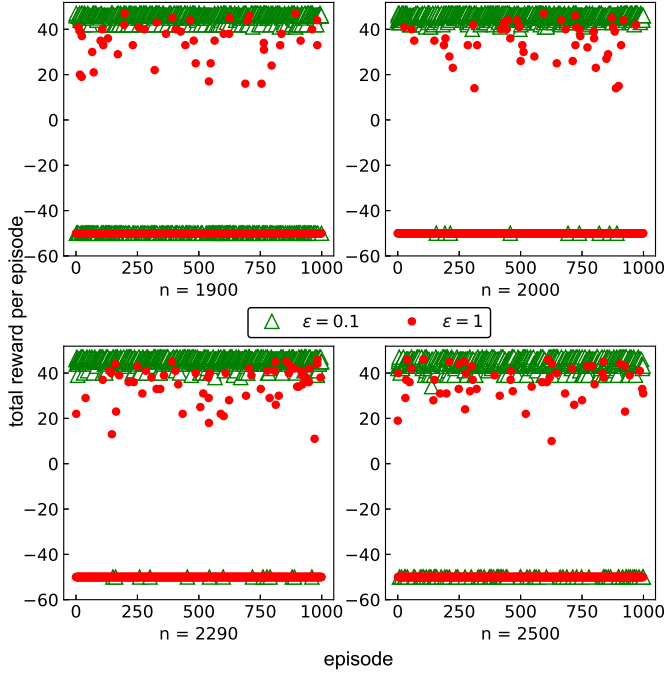}
       \caption{Comparison of the agent's copies ($\epsilon = 0.01$) at different training steps during the learning process shown in figure \ref{fig:PTprogress} to untrained ones ($\epsilon = 1$). The trained agents receive more rewards per episode than the untrained ones.}
    \label{fig:PTcomparison_other}
    
\end{figure}

For the analysis of a trained agent's behaviour, we extract the agent's copy at the 1900th, 2000th, 2290th and the 2500th training step where its performance has local maxima. Its behaviour at these trainining steps is compared to the performance of an untrained agent in figure \ref{fig:PTcomparison_other}.

 Table \ref{tab:PTsequxp} and table \ref{tab:PTsequxm} list the shortest successful sequences for $\ket{x+}$ and $\ket{x-}$ as the central spin starting states the trained agents acting with $\epsilon = 0.01$ can find.
\begin{table*} 
\caption{Shortest sequences of time evolutions and subsequent projective measurements found by the agent's copies at different training steps that lead to the $\ket{\Psi^-}$ state for the $\ket{x+}$ starting state. As opposed to before, this agent was trained using the Double DQN algorithm.}
\centering
\begin{tabular}{|c|c|c|c|}
\hline
  training step & sequence &fidelity&success rate \\ \hline
1900 &  unsuccessful & &  \\\hline 
2000 & --$U(\tau)$-$P_{x+}$-$U(\tau)$-$P_{x+}$-$U(\tau)$-$P_{x+}$-$U(\tau)$-$P_{x+}$-- & 99.227 & 25.391\\\hline 
2290 &  --$U(\tau)$-$P_{y-}$-$U(\tau)$-$P_{y-}$-$U(\tau)$-$P_{y-}$-$U(\tau)$-$P_{y-}$-$U(\tau)$-$P_{y-}$-- & 99.227 & 12.695 \\\hline 
2500 & --$U(\tau)$-$P_{y-}$-$U(\tau)$-$P_{y-}$-$U(\tau)$-$P_{y-}$-$U(\tau)$-$P_{x+}$-$U(\tau)$-$P_{x+}$--& 99.227 & 6.348 \\\hline \end{tabular}

\label{tab:PTsequxp}
\end{table*}

\begin{table*} 
\centering
\caption{Shortest sequences of time evolutions and subsequent projective measurements that lead to the $\ket{\Psi^-}$ state for the $\ket{x-}$ starting state found by the Double DQN agent's copies at different training steps.}
\begin{tabular}{|c|c|c|c|}
\hline
  training step & sequence &fidelity&success rate \\ \hline
1900 & --$U(\tau)$-$P_{x-}$-$U(\tau)$-$P_{x-}$-$U(\tau)$-$P_{x-}$-$U(\tau)$-$P_{x-}$-- & 99.227 & 25.391   \\\hline 
2000 &  unsuccessful &  &  \\\hline 
2290 &  --$U(\tau)$-$P_{y-}$-$U(\tau)$-$P_{y-}$-$U(\tau)$-$P_{y-}$-$U(\tau)$-$P_{y-}$-$U(\tau)$-$P_{y-}$--& 99.227 & 12.695 \\\hline 
2500 &  --$U(\tau)$-$P_{y-}$-$U(\tau)$-$P_{y-}$-$U(\tau)$-$P_{y-}$-$U(\tau)$-$P_{x+}$-$U(\tau)$-$P_{x+}$--&99.227 & 6.348 \\\hline \end{tabular}

\label{tab:PTsequxm}
\end{table*}

\begin{figure}[t!]
    \centering

        \includegraphics[width=0.75\textwidth]{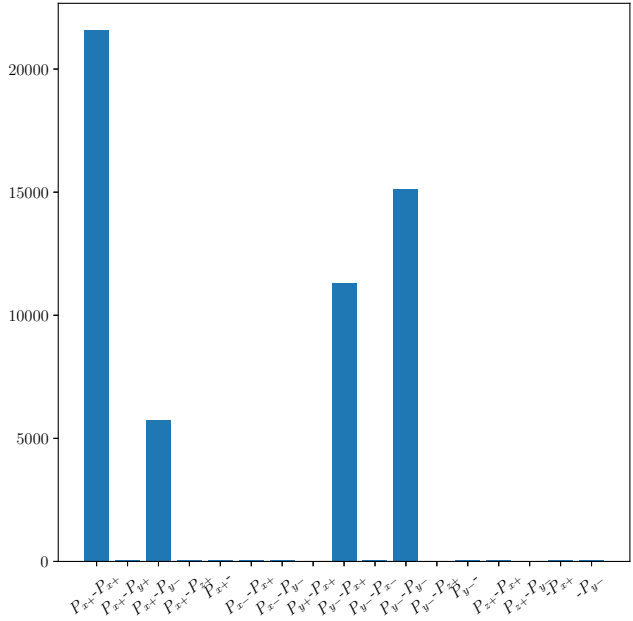}
       \caption{Counts of appearances of action combinations in 10000 sequences leading to $\ket{\Psi^-}$ with a random starting state the agent's copy extracted at the 2500th training step has found. The agent follows an $\epsilon$-greedy policy with $\epsilon = 0.01$. The dash stands for 'do nothing'. The action combinations with insignificant appearance counts have been omitted from the histogram. All sequences overwhelmingly consist of combinations of projections onto the superposition states $\ket{x\pm}$ and $\ket{y\pm}$.}
    \label{fig:PTHist2500}
    
\end{figure}

Together with the histogram in figure \ref{fig:PTHist2500} they show that for random  or the $\ket{x\pm}$ starting states, the agent prefers repeating projective measurements onto a single subspace which belongs to $\ket{x+}$,$\ket{x-}$, $\ket{y+}$ or $\ket{y-}$. Apart from that, it also mixes projective measurements onto these subspaces. We therefore conclude that in order for the two nuclear spins to end up in the $\ket{\Psi^-}$ state, one has to repeatedly perform projections onto one of the superposition subspaces, no matter what the state of the central spin was in the beginning. Hence, with the help of deep reinforcement learning we have generalised the findings of \citep{greiner_purification_2017} for the case of two nuclear spins.

\section{\label{sec: conclusion}Conclusion and Discussion}

We applied the technique of Deep Reinforcement Learning to the task of Quantum state engineering. An agent taking the role of the experimentalist had to find sequences for the time evolution of a central spin system consisting of an NVC electron spin coupled to a nuclear spin bath followed by projective measurements of the electron spin in order to polarise said spin bath in a desired way. 
 
 We looked at the $m_{\mathrm{s}}=\pm 1$ subspace of the central spin and two $\ce{^{13}C}$ spins to which it is coupled to and instructed agents to find sequences that take the nuclear spins from a completely mixed state to the maximally entangled Bell states. The agent did indeed find sequences composed of projective measurements and free evolution periods that lead to the formation of any of the four Bell states. For the formation of singlet state, $\ket{\Psi^-}$ the agent found an identical sequence as Greiner et. al in \citep{greiner_purification_2017}. However, it also generalised these human found solutions by finding many more such sequences that lead to the formation of a similar entangled state.
 
As the entire density matrix was included into the state representation and the reward was defined as given in equation \eqref{eq:reward}, the transition probability of the environment and the probability of the agent receiving a reward $r$ only depended on the current state $s$ and selected action $a$. Therefore, the problem constituted a Markov Decision Process allowing it to be studied under the framework of RL. Analogously, similar quantum control tasks could be translated to the RL paradigm, given that there exists some entity as the density matrix containing all information about the system's subsequent dynamics independent of its history.  \\

The implementation of a Deep Reinforcement Learning algorithm for a larger spin ensemble is straight-forward and could theoretically lead to more sophisticated quantum protocols for the task of creating highly entangled quantum states in a large nuclear spin bath. However, as the number of entries of the density matrix describing the entire system grows exponentially with the number of spins, the DQN and DDQN with a state representation as the one we have chosen, will become very difficult to scale. (For RL to work the environment's dynamics have to be constantly simulated all over again. In our case, this is what took the most time.) For this reason, we suggest to study this further by reducing the dimension of the state representation and formulating the problem as a partially observable Markov Decision Process (POMDP), where the state representations are replaced by observations that do not fulfil the Markov property, which could then be handled by a recurrent neural network. \citep{hausknecht_deep_2015} 

 After reducing the state representations to observations and thereby improving the scalability of the algorithm one could then try to implement the RL learning protocol in real life, i.e. let the agent perform its actions on an actual NV-centre and optimise its policy for observables accessible during an experiment. The costliest part of our algorithm, the simulation of the quantum dynamics would therefore be outsourced to a real quantum system. Also, in the future, when there may be devices suited to simulate the dynamics of a large spin ensemble,the task of computing the system dynamics could be managed by such a device.  \\
 
 Another possibility would be to use all previous actions as the state representations, i.e. the agent would change the environment's state from "$P_{x+}-P_{x-}$" to "$P_{x+}-P_{x-}-P_{y+}$" by selecting the action $P_{y+}$, and implement the projective simulation algorithm proposed by Hans J. Briegel and Gemma De las Cuevas \citep{briegel_projective_2012}. Projective simulation has been used to create photonic quantum experiments where the agent receives the entire optical setup as the state representation and changes the state of the environment by adding an optical element to the setup \citep{melnikov_active_2018}. This would be analogous to a problem where the agent takes in the entire history of projections and time evolution periods as the state representation and changes the state by adding another action, i.e. another projection (or none). 
 
 To conclude: We have theoretically described how measurement-induced Quantum State Engineering could be achieved within the framework of RL. We found that the agent could find optimal measurement sequences that lead the system towards a desired entangled state.  In general, RL could be applied for other quantum protocols if they have some sort of trial-and-error nature. However, combining machine learning on a classical computer with problems including a quantum system becomes increasingly difficult due to an exponential growth in the parameter space. Only upon reducing/optimizing the information flow to the agent  and further replacing the classical computation of quantum dynamics with a quantum simulation (given that the quantum simulator is powerful enough) or by an actual experimental setup, one would be able to harness the abilities of RL in quantum information processing.


%

\begin{acknowledgements}

The authors acknowledge support from the Max Planck Society as well as the EU via the project ASTERIQS and QIA, the ERC grant SMeL and the DFG (FOR 2724). 
 
\end{acknowledgements}

%
\section*{Conflict of interest}
 The authors declare that they have no conflict of interest.


\bibliographystyle{spbasic}

\bibliography{bibref}

\end{document}